\documentclass[aps,prl,preprint,tightenlines,superscriptaddress,showpacs,byrevtex]{revtex4}
\usepackage{graphicx} 
\usepackage{epsfig}   
\usepackage{dcolumn}  

\graphicspath{{ps}}

\begin{document}


\preprint{\vbox{ \hbox{   }
                 \hbox{BELLE-CONF-0427}
                 \hbox{ICHEP04 8-0674} 
}}

\title{ \quad\\[0.5cm] Measurement of Exclusive $B \to X_u \ell \nu$ Decays \\
with $D^{(*)} \ell \nu$ Decay Tagging}


\affiliation{Aomori University, Aomori}
\affiliation{Budker Institute of Nuclear Physics, Novosibirsk}
\affiliation{Chiba University, Chiba}
\affiliation{Chonnam National University, Kwangju}
\affiliation{Chuo University, Tokyo}
\affiliation{University of Cincinnati, Cincinnati, Ohio 45221}
\affiliation{University of Frankfurt, Frankfurt}
\affiliation{Gyeongsang National University, Chinju}
\affiliation{University of Hawaii, Honolulu, Hawaii 96822}
\affiliation{High Energy Accelerator Research Organization (KEK), Tsukuba}
\affiliation{Hiroshima Institute of Technology, Hiroshima}
\affiliation{Institute of High Energy Physics, Chinese Academy of Sciences, Beijing}
\affiliation{Institute of High Energy Physics, Vienna}
\affiliation{Institute for Theoretical and Experimental Physics, Moscow}
\affiliation{J. Stefan Institute, Ljubljana}
\affiliation{Kanagawa University, Yokohama}
\affiliation{Korea University, Seoul}
\affiliation{Kyoto University, Kyoto}
\affiliation{Kyungpook National University, Taegu}
\affiliation{Swiss Federal Institute of Technology of Lausanne, EPFL, Lausanne}
\affiliation{University of Ljubljana, Ljubljana}
\affiliation{University of Maribor, Maribor}
\affiliation{University of Melbourne, Victoria}
\affiliation{Nagoya University, Nagoya}
\affiliation{Nara Women's University, Nara}
\affiliation{National Central University, Chung-li}
\affiliation{National Kaohsiung Normal University, Kaohsiung}
\affiliation{National United University, Miao Li}
\affiliation{Department of Physics, National Taiwan University, Taipei}
\affiliation{H. Niewodniczanski Institute of Nuclear Physics, Krakow}
\affiliation{Nihon Dental College, Niigata}
\affiliation{Niigata University, Niigata}
\affiliation{Osaka City University, Osaka}
\affiliation{Osaka University, Osaka}
\affiliation{Panjab University, Chandigarh}
\affiliation{Peking University, Beijing}
\affiliation{Princeton University, Princeton, New Jersey 08545}
\affiliation{RIKEN BNL Research Center, Upton, New York 11973}
\affiliation{Saga University, Saga}
\affiliation{University of Science and Technology of China, Hefei}
\affiliation{Seoul National University, Seoul}
\affiliation{Sungkyunkwan University, Suwon}
\affiliation{University of Sydney, Sydney NSW}
\affiliation{Tata Institute of Fundamental Research, Bombay}
\affiliation{Toho University, Funabashi}
\affiliation{Tohoku Gakuin University, Tagajo}
\affiliation{Tohoku University, Sendai}
\affiliation{Department of Physics, University of Tokyo, Tokyo}
\affiliation{Tokyo Institute of Technology, Tokyo}
\affiliation{Tokyo Metropolitan University, Tokyo}
\affiliation{Tokyo University of Agriculture and Technology, Tokyo}
\affiliation{Toyama National College of Maritime Technology, Toyama}
\affiliation{University of Tsukuba, Tsukuba}
\affiliation{Utkal University, Bhubaneswer}
\affiliation{Virginia Polytechnic Institute and State University, Blacksburg, Virginia 24061}
\affiliation{Yonsei University, Seoul}
  \author{K.~Abe}\affiliation{High Energy Accelerator Research Organization (KEK), Tsukuba} 
  \author{K.~Abe}\affiliation{Tohoku Gakuin University, Tagajo} 
  \author{N.~Abe}\affiliation{Tokyo Institute of Technology, Tokyo} 
  \author{I.~Adachi}\affiliation{High Energy Accelerator Research Organization (KEK), Tsukuba} 
  \author{H.~Aihara}\affiliation{Department of Physics, University of Tokyo, Tokyo} 
  \author{M.~Akatsu}\affiliation{Nagoya University, Nagoya} 
  \author{Y.~Asano}\affiliation{University of Tsukuba, Tsukuba} 
  \author{T.~Aso}\affiliation{Toyama National College of Maritime Technology, Toyama} 
  \author{V.~Aulchenko}\affiliation{Budker Institute of Nuclear Physics, Novosibirsk} 
  \author{T.~Aushev}\affiliation{Institute for Theoretical and Experimental Physics, Moscow} 
% ...
\collaboration{The Belle Collaboration}

\begin{abstract}
We report on a measurement of the charmless semileptonic $B$ decays, 
$B^0 \to \pi^- \ell^{+} \nu$ and 
$B^0 \to \rho^- \ell^{+} \nu$, 
based on 140 fb$^{-1}$ data collected with the Belle detector at the KEKB 
$e^+e^-$ asymmetric collider.
In this analysis, accompanying $B$ mesons are reconstructed from
semileptonic $B \rightarrow D^{(*)} \ell \nu$ decays, which enables us to detect the signal with high purity 
and with marginal statistics.
We found branching fractions of 
${\cal B}(B^0 \to \pi^- \ell^+ \nu)  = 
(1.76 \pm 0.28 \pm 0.20 \pm 0.03) \times 10^{-4}$ and
${\cal B}(B^0 \to \rho^- \ell^+ \nu)  = 
(2.54 \pm 0.78 \pm 0.85 \pm 0.30) \times 10^{-4}$,
where the errors are statistical, experimental systematic, systematic due
to form-factor uncertainties.
We present also the branching fractions in three $q^2$ intervals; 
$q^2 < 8$, $8 \leq q^2 < 16$, $q^2 \geq 16$\,GeV$^2/c^2$, 
for each decay mode.
Based on these results, the magnitude of the Cabibbo-Kobayashi-Maskawa matrix 
element $V_{ub}$ is extracted. 
All of the presented results are preliminary.
\end{abstract}


\maketitle

\tighten

{\renewcommand{\thefootnote}{\fnsymbol{footnote}}}
\setcounter{footnote}{0}

\section{Introduction}
\label{sec:Introduction}
Exclusive $B \to X_u \ell \nu$ decays proceed via $b \to u W^-$ tree 
diagram with a spectator quark, and can be used to determine $|V_{ub}|$, 
one of the smallest and least known elements of the 
Cabibbo-Kobayashi-Maskawa matrix~\cite{KM}.
The major theoretical issue is determination of the form-factors (FF)
involved in the decays. 

In principle lattice QCD(LQCD) provides the most reliable calculation of
FFs, but so far only quenched calculations have been available.
Recently preliminary results of unquenched calculations have become reported
~\cite{Okamoto,Shigemitsu} so that a model-independent determination of $|V_{ub}|$
becomes feasible. Since LQCD results are available only in the high $q^2$ region
the clean extraction of the $B \to \pi \ell \nu$ branching fraction in the high $q^2$ region
($\geq 16$\,GeV$^2/c^2$) is essential.


There have been several measurements in the past by CLEO, BaBar and
Belle for the $B \to \pi \ell \nu$, $\rho \ell \nu$, $\eta \ell \nu$
and $\omega \ell \nu$ modes
~\cite{CLEO1996,CLEO2000,CLEO2003,BABAR2003,BELLE2003}.
All of these analyses utilize the method, originally 
developed by CLEO, where the $B$ decays are reconstructed by inferring 
the undetected neutrino momentum from the missing energy and momentum
(``$\nu$-reconstruction method'')~\cite{CLEO1996}.
In the $B$-factory era, we may quickly improve the statistical 
precision by simply applying the $\nu$-reconstruction method.
However, the systematic uncertainty may soon limit the experimental 
uncertainty because of the poor signal-to-noise ratio.




In this paper, we present measurements of $B^0 \to \pi^- \ell^+ \nu$
and $B^0 \to \rho^- \ell^+ \nu$ decays by $D^{(*)} \ell \nu$ decay tagging,
where we reconstruct the whole decay chain from the $\Upsilon(4S)$;
$\Upsilon(4S) \to B_{sig}B_{tag}$, $B_{sig} \to \pi^-(\rho^-) \ell^+ \nu$ and
$B_{tag} \to D^{(*)+} \ell^- \bar{\nu}$ with several $D^{(*)+}$ sub
decay modes.
The back-to-back correlation of the two $B$ mesons in the $\Upsilon(4S)$
rest frame allows us to constrain the kinematics of the double
semileptonic decay.
Belle has presented a preliminary measurement of the inclusive
$B \to X_u \ell \nu$ decay using this method~\cite{BELLE_Sugiyama}. 
While this method gives a signal detection efficiency that is significantly 
lower than the $\nu$-reconstruction method because of the tagging,
it allows us to detect the signal with high purity, even with a relaxed
cut on the lepton momentum down to 0.8\,GeV/$c$.
This leads to less experimental systematic uncertainty and promises to yield the best 
overall precision from the large data sample being accumulated by $B$-factories over the next few years.
We present the simultaneous extraction of $B^0 \to \pi^- \ell^+ \nu$ 
and $B \to \rho^- \ell^+ \nu$ and their $q^2$ distribution.
Inclusion of the charge conjugate decays is implied throughout this paper.

\section{Data Set and Experiment}
\label{sec:data_exp}

The analysis is based on data recorded with the Belle detector at the 
asymmetric $e^+e^-$ collider KEKB operating at the center-of-mass (c.m.) 
energy of the $\Upsilon(4S)$ resonance~\cite{KEKB}.
The $\Upsilon(4S)$ data set used for this study corresponds to an
integrated luminosity of 140 fb$^{-1}$ and contains $152 \times 10^6$ 
$B \bar{B}$ events.

The Belle detector is a large-solid-angle magnetic spectrometer
that consists of a three-layer silicon vertex detector (SVD),
a 50-layer central drift chamber (CDC), 
an array of aerogel threshold \v{C}erenkov counters (ACC),
a barrel-like arrangement of time-of-flight scintillation counters (TOF),
and an electromagnetic calorimeter comprised of CsI(Tl) crystals (ECL)
located inside a super-conducting solenoid coil 
that provides a 1.5~T magnetic field.  
An iron flux-return located outside of the coil is instrumented
to detect $K_L^0$ mesons and to identify muons (KLM).  
The detector is described in detail elsewhere~\cite{BELLE}. 

A detailed Monte Carlo (MC) simulation, which fully describes the detector
geometry and response and is based on GEANT~\cite{GEANT}, is
applied to estimate the signal detection efficiency and to study the 
background.
To examine the FF dependence, MC samples for the 
$B^0 \to \pi^-(\rho^-) \ell^{+} \nu$ signal decays are generated 
with different form-factor models;
a quark model (ISGW~II ~\cite{ISGW2}), 
light cone sum rules (LCSR; Ball'01 for $\pi^- \ell^+ \nu$ ~\cite{Ball01} 
and Ball'98 for $\rho^- \ell^+ \nu$ ~\cite{Ball98}) and 
a quenched lattice QCD (UKQCD ~\cite{UKQCD}).
A relativistic quark model (Melikhov ~\cite{Melikhov}) is also 
used for $\rho^- \ell^+ \nu$.
To model the cross-feed from other $B \to X_u \ell \nu$ decays,
MC samples are generated with the ISGW~II model for the resonant 
components and the DeFazio-Neubert model ~\cite{Fazio-Neubert} for 
non-resonant component ($\pi \ell \nu$ and $\rho \ell \nu$ 
components are excluded in this sample).
A MC sample for the $B^0 \to D^{*-} \ell^+ \nu$ decay, to be used
for calibration as described later, is also prepared.  
To model the $B\bar{B}$ and continuum backgrounds, large generic 
$B\bar{B}$ (based on QQ98~\cite{QQ98}) and $q\bar{q}$ Monte Carlo 
samples are used.


\section{Event Reconstruction and Selection}

Charged particle tracks are reconstructed from hits in the SVD and CDC. 
They are required to satisfy track quality cuts based on their impact 
parameters relative to the measured profile of the interaction point 
(IP profile) of the two beams. 
Charged kaons are identified by combining information on ionization loss 
($dE/dx$) in the CDC, \v{C}herenkov light yields in the ACC and time-of-flight 
measured by the TOF system.
For the nominal requirement, the kaon identification efficiency is 
approximately $88\%$ and the rate for misidentification of pions as 
kaons is about $8\%$.
Hadron tracks that are not identified as kaons are treated as pions.
Tracks satisfying the lepton identification criteria, as described later, 
are removed from consideration.

Neutral pions are reconstructed using $\gamma$ pairs with an invariant mass 
between 117 and 150\,MeV/$c^2$.
Each $\gamma$ is required to have a minimum energy deposit of 
$E_{\gamma} \geq 50$\,MeV.
$K_S^0$ mesons are reconstructed using pairs of charged tracks that
have an invariant mass within $\pm 7.6$\,MeV/$c^2$ of the known $K_S^0$ 
mass.

Electron identification is based on a combination of $dE/dx$ in CDC,
the response of ACC, shower shape in ECL and the ratio of energy 
deposit in ECL to the momentum measured by the tracking system.
Muon identification by KLM is performed by resistive plate counters 
interleaved in the iron yoke.
The lepton identification efficiencies are estimated to be about 90\% 
for both electrons and muons in the momentum region above 1.2\,GeV/$c$. 
where leptons from the prompt $B$ decays dominate.
The hadron misidentification rate is measured 
using reconstructed $K_S^0 \to \pi^+ \pi^-$
and found to be less than 0.2\% for electrons and 1.5\% for muons
in the same momentum region.

For the reconstruction of $B_{tag} \to D^{(*)+} \ell^- \bar{\nu}$, 
the lepton candidate is required to have the right sign charge with
respect to the $D$ meson flavor and the laboratory momentum greater
than 1.0\,GeV/$c$ ($p_{\ell}^{lab} > 1.0$ GeV/$c$). 
The $D^{(*)+}$ candidates are reconstructed by using four decay modes of $D^+$ -- 
$D^+ \to K^- \pi^+ \pi^+$, $K^- \pi^+ \pi^+ \pi^0$, $K_S^0 \pi^+$ and 
$K_S^0 \pi^+ \pi^0$ -- and six decay modes of $D^0$ -- 
$K^- \pi^+$, $K^- \pi^+ \pi^0$, $K^- \pi^+ \pi^+ \pi^-$, $K_S^0 \pi^0$, 
$K_S^0 \pi^+ \pi^-$, $K_S^0 \pi^+ \pi^- \pi^0$.
The candidates are required to have an invariant mass $m_D$ within
$\pm 2\sigma$ ($\sigma$ is a standard deviation) of the nominal $D$ mass, 
where the mass resolution $\sigma$ is dependent on the decay mode. 
$D^{*+}$ mesons are reconstructed by combining the $D^0$ candidate and a pion, 
$D^{*+} \to D^0 \pi^+$ and $D^{*+} \to D^+ \pi^0$.
They are required to have the mass difference $\Delta m = m_{\bar{D}\pi} - m_{\bar{D}}$ 
within $\pm 2 \sigma$ of the nominal values.    


 For the reconstruction of $B_{sig} \to X_u^- \ell^+ \nu$, 
the lepton candidate is required to have the right sign charge with
respect to the $X_u$ system and $p_{\ell}^{lab} > 0.8$ GeV/$c$.
The $X_u$ system may consist of either a single charged pion or a charged
pion and neutral pion candidate, and the event may contain no additional
charged tracks or $\pi^0$ candidates ($N_{\pi^-} = 1$ and $N_{\pi^0} \leq
1$). We also require that the residual energy from neutral clusters is
less than 0.3 GeV ($E_{neut} < 0.3$\,GeV).   
The two leptons on the tag and the signal sides are required to have opposite charge.
Loss of the signal due to $B^0 - \bar{B^0}$ mixing is estimated by
the MC simulation.

\begin{figure}[htbp]
 \begin{center}
  \mbox{\psfig{figure=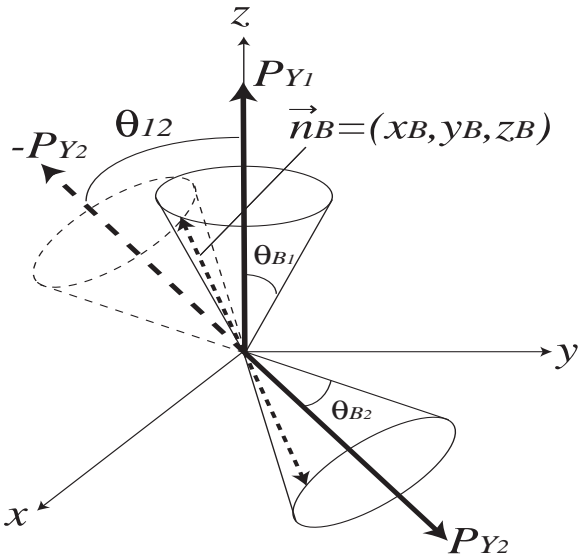,width=2.5in, height=2.5in, angle=0, scale=1.2 } } 
  \caption{Kinematics of the double semileptonic decay.}
  \label{fig:double_cone}
  \end{center}
\end{figure}

We then impose a constraint from kinematics of the double semileptonic decay
in the $\Upsilon(4S)$ rest frame.
In the semileptonic decay on each side, $B_{1(2)} \to Y_{1(2)} \nu_{1(2)}$ 
($Y_1 = D^{(*)+} \ell^-$ and $Y_2 = X_u \ell^+$), the angle between the 
$B_{1(2)}$ meson and the detected $Y_{1(2)}$ system 
$\theta_{B_{1(2)}}$ is calculated from the relation,
$p_{\nu}^2 = (p_{B} - p_{Y})^2 = 0$ and the known $P_B$ (the absolute momentum 
of the mother $B$ meson).
This means that the $B_{1(2)}$ direction is constrained on the surface of a 
cone defined with the angle $\theta_{B_{1(2)}}$ around the direction 
of the $Y_{1(2)}$ system, as shown graphically in 
Figure~\ref{fig:double_cone}.
Then the back-to-back relation of the two $B$ meson directions implies that
the real $B$ direction is on the cross lines of the two cones when one of
the $B$ system is spatially inverted.
Denoting $\theta_{12}$ the angle between the $D^{*+} \ell^-$ and the 
$X_u \ell^+$ systems, the $B$ directional vector $\vec{n}_B = (x_B, y_B, z_B)$ 
is given by, 
$z_B = \mbox{cos}\theta_{B_1}$,  
$y_B = (\mbox{cos}\theta_{B_2} - \mbox{cos}\theta_{B_2}\mbox{cos}\theta_{12})
     / \mbox{sin}\theta_{12}$,
and

\begin{eqnarray}
 x_B = \pm \sqrt{1-\frac{1}{\mbox{sin}\theta_{12}}
(\mbox{cos}^2\theta_{B_1} + \mbox{cos}^2\theta_{B_2} 
- 2 \mbox{cos}\theta_{B_1}\mbox{cos}\theta_{B_2}\mbox{cos}\theta_{12})}
 \label{eq:x_B}
\end{eqnarray}
with the coordinate definition in Figure~\ref{fig:double_cone}.
If the hypothesis of the double semileptonic decay is correct and all
the decay products are detected except for the two neutrinos, $x_B^2$
must range from 0 to 1.
Events passing a rather loose cut $x_B^2 > -2.0$ are used for the signal 
extraction in the later stage of the analysis.

In this analysis, Eq.~\ref{eq:x_B} has two solutions
and the direction of the $B$ meson is not uniquely determined.
We calculate, therefore, $q^2$ as $q^{2} = (E^{*}_{beam} - E^{*}_{X_u})^2 - p^{*}_{X_u}{^2}$,
using the beam energy ($E^*_{beam}$), energy ($E^*_{X_u}$) and 
momentum ($p^*_{X_u}$) of the $X_u$ system and
neglecting the momentum of the $B$ meson in the c.m. system.
The signal Monte Carlo simulation predicts that the average $q^2$ resolution
is approximately 0.75\,GeV$^2/c^2$.

According to Monte Carlo simulations, the major backgrounds
come from $B^0\bar{B^0}$ events and other $B^0 \to X_u^- \ell^+ \nu$
decays, where some particles escape detection.
Other backgrounds from $B^+ B^-$ events, $B^+ \to X_u^0 \ell^+ \nu$ decays
and $q \bar{q}$ processes are small.

With the event selection described above, the signal MC simulation 
indicates that the total detection efficiency ($\epsilon_{total}$), on the 
average of the electron and muon channels, is
$1.5 \times 10^{-3}$ for $\pi^- \ell^+ \nu$ and
$8.1 \times 10^{-4}$ for $\rho^- \ell^+ \nu$
with the LCSR model (Ball'01 for $\pi^- \ell^+ \nu$ and Ball'98 for
$\rho^- \ell^+ \nu$).
Here, $\epsilon_{total}$ is defined with respect to the number of 
produced $B^0 \bar{B^0}$ pairs, where one $B$ decays into the signal mode, and 
includes loss of the signal due to $B^0 - \bar{B^0}$ mixing.
Because of the relaxed lepton momentum cut ($>0.8$\,GeV/$c$), the variation 
of the efficiency with different FF models is relatively small;
the maximum deviation from LCSR is $-1.4$\% with ISGW~II for 
$\pi^- \ell^+ \nu$ and $+5.5$\% with UKQCD for $\rho^- \ell^+ \nu$.

\begin{figure}[htbp]
\vspace{1.0cm}
  \begin{center}
    \mbox{\psfig{figure=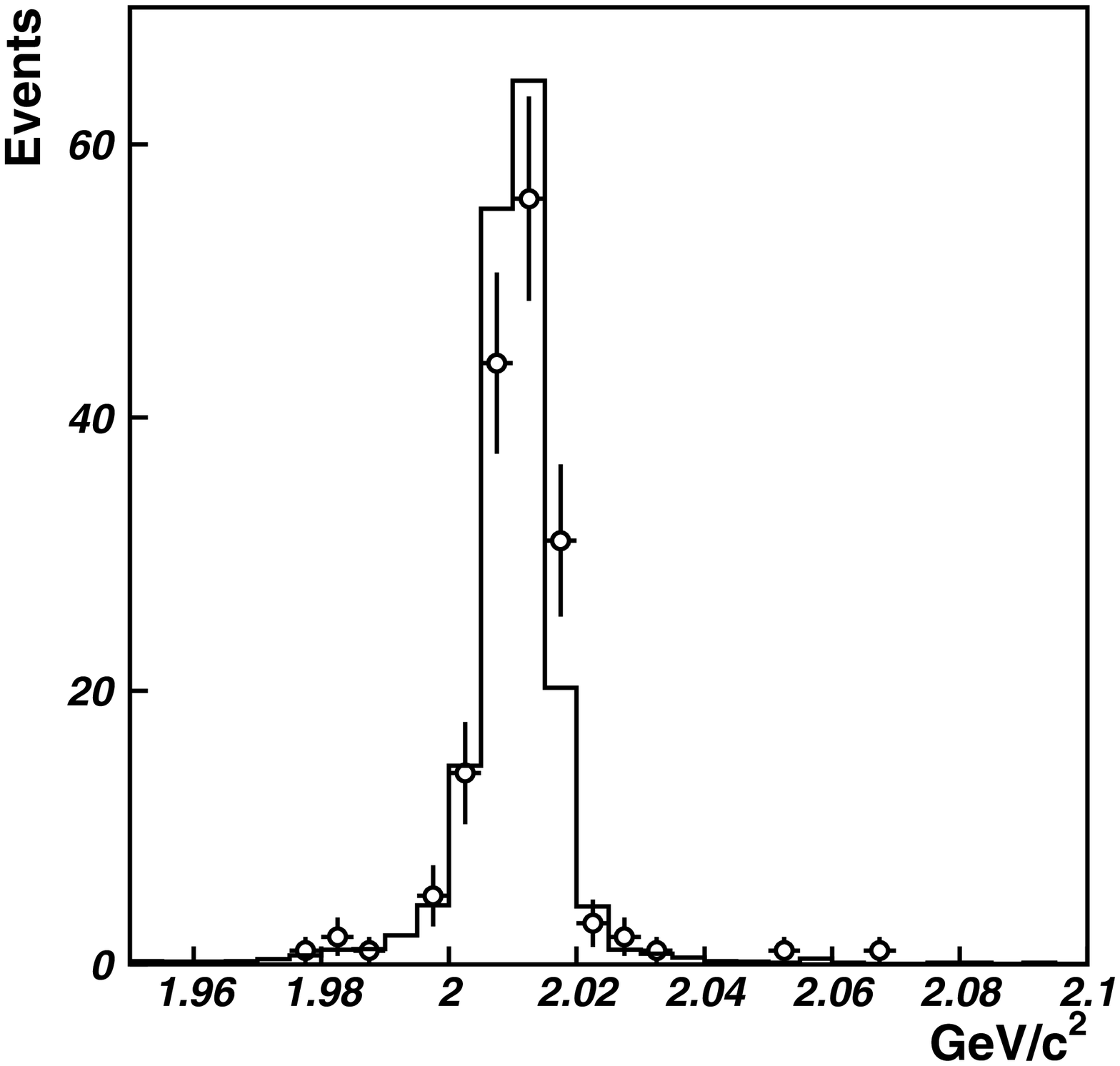,width=2.5in, height=2.5in, angle=0, scale=1. } }
    \mbox{\psfig{figure=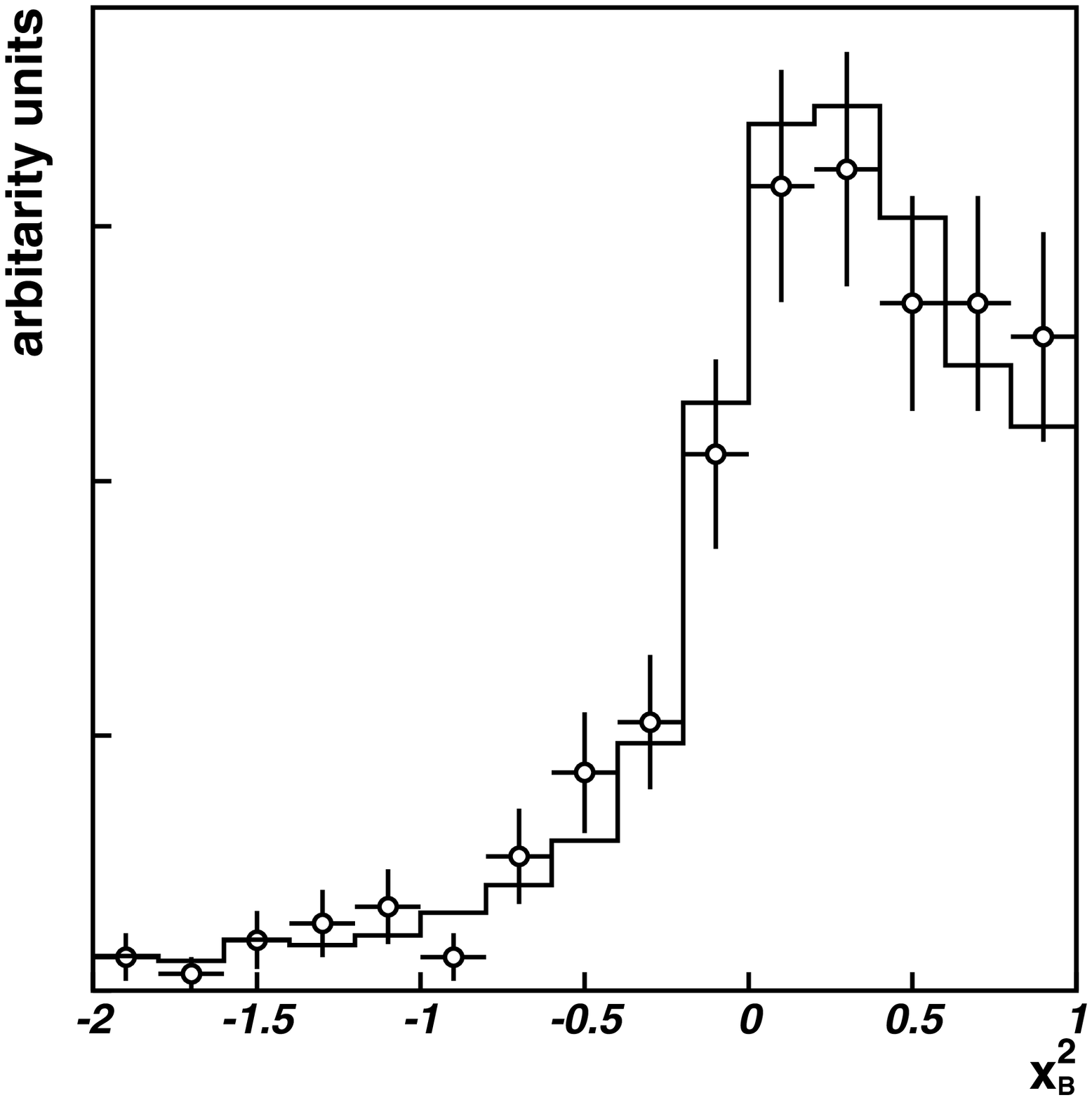,width=2.5in, height=2.5in, angle=0, scale=1. } }
    \caption{ Reconstructed $M(K\pi\pi)$ distribution(left) and $x^2_B$ distribution(right) 
   for the $D^{*-} \ell^{+} \nu$ calibration decay. Open circle is data and histogram is signal MC.}
    \label{fig:dstlnu}
  \end{center}
\end{figure}

The validity of the method to reconstruct the double semileptonic decay 
is checked by using the decay, $B_{sig}^0 \to D^{*-} \ell^{+} \nu$ followed 
by $D^{*-} \to \bar{D^0} \pi^{-}, \bar{D^0} \to K^{+} \pi^{-}$, with 
the same requirement on the tagging side.
Figure~\ref{fig:dstlnu}-a) shows the obtained $M_{K\pi\pi}$ distribution
and its comparison to the MC expectation.
With the 140 fb$^{-1}$ data sample, $147 \pm 12$ decays are clearly 
identified, while $165 \pm 9$ events are expected based on the product branching
fraction ${\cal B}(B^0 \to D^{*-} \ell^{+} \nu, D^{*-} \to \bar{D^0} \pi^{-},
\bar{D^0} \to K^{+} \pi^{-}) = (1.40 \pm 0.07) \times 10^{-3}$ deduced
from ~\cite{PDG2004}.
Their ratio $R = 0.89 \pm 0.08$ is used to correct the total detection
efficiency $\epsilon_{total}$ for the $\pi(\rho) \ell \nu$ signals 
predicted by the signal MC simulations.
Figure~\ref{fig:dstlnu}-b) shows the comparison of the reconstructed 
$x_B^2$ distribution using the same sample to the MC simulation.
The agreement between the data and MC demonstrates the validity of the
present measurement.

\section{Extraction of Branching Fractions}
\label{sec:SignalExtraction}
The $B^0 \to \pi^- \ell^+ \nu$ and $B^0 \to \rho^- \ell^+ \nu$
signals are extracted by fitting the obtained two-dimensional distribution 
in $(x_B^2, M_X)$, where $M_X$ is the invariant mass of the $X_u$ system.
Here, the fit components are the two signal modes; $B^0 \to \pi^- \ell^+ \nu/ 
\rho^- \ell^+ \nu$, the other $B^0 \to X_u^- \ell^+ \nu$ background and the 
$B \bar{B}$ background.
The PDF (probability distribution function) for each component is deduced
from the MC simulation. 
The $\pi^-(\rho^-) \ell^+ \nu$ signal events exhibit characteristic behavior in 
both of their $x_B^2$ and $M_X$ distributions; other $B^0 \to X_u^- \ell^+ \nu$ events exhibit a weak
peaking structure in $x_B^2$ but a relatively flat distribution in $M_X$;
the $B \bar{B}$ background events are distributed uniformly in both variables. 
 The total inclusive branching fraction ${\cal B}(B \to X_u \ell \nu)$ 
is constrained to be 0.25$\%$~\cite{BELLE_Kakuno}.

\begin{figure}[htbp]
 \begin{center}
  \begin{tabular}{ccc}
   \hspace{-0.8cm}
   \mbox{\psfig{figure=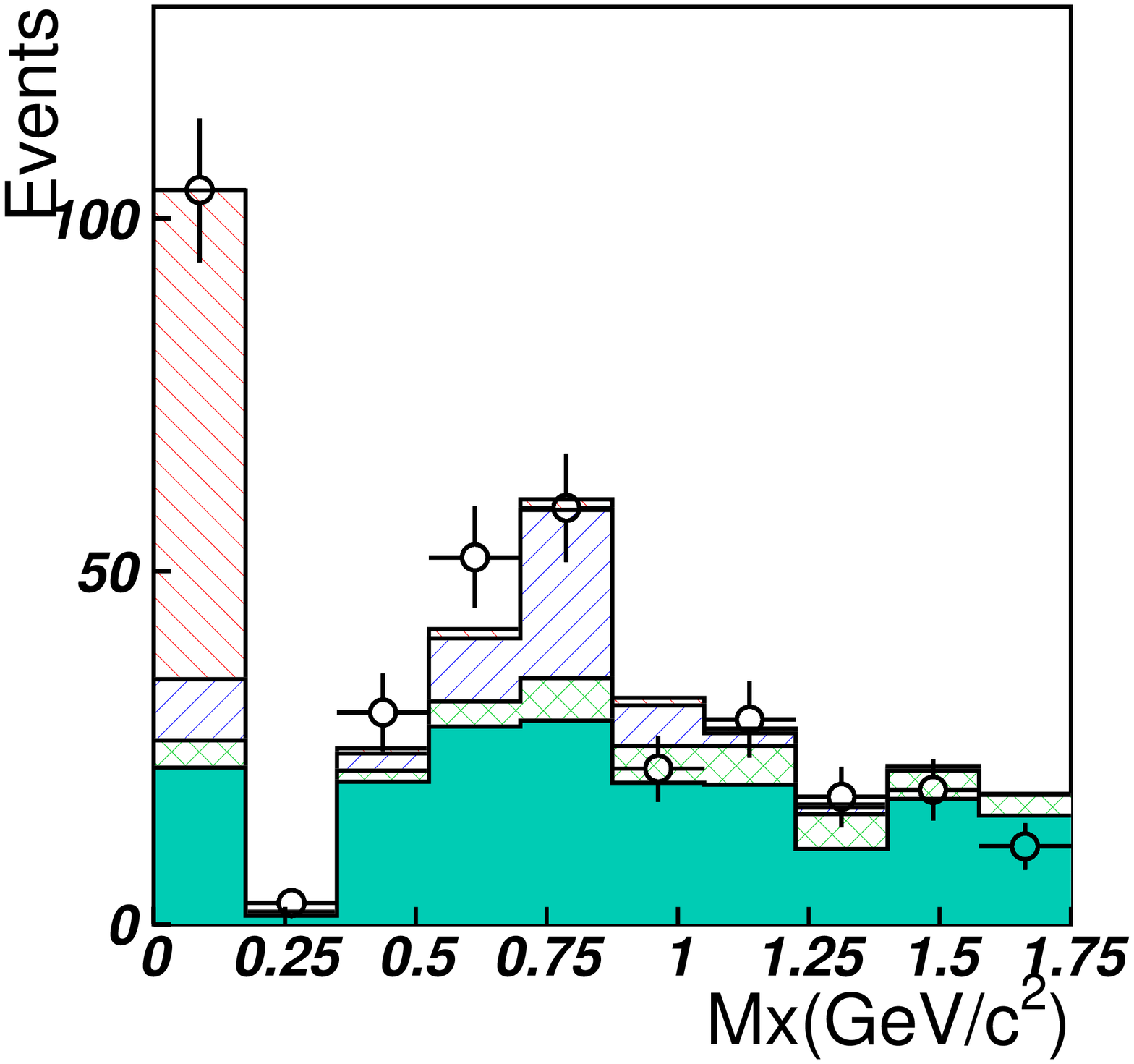,width=2.3in, height=2.3in, angle=0, scale=1.0 } } 
   \hspace{-0.8cm}{\mbox{\psfig{figure=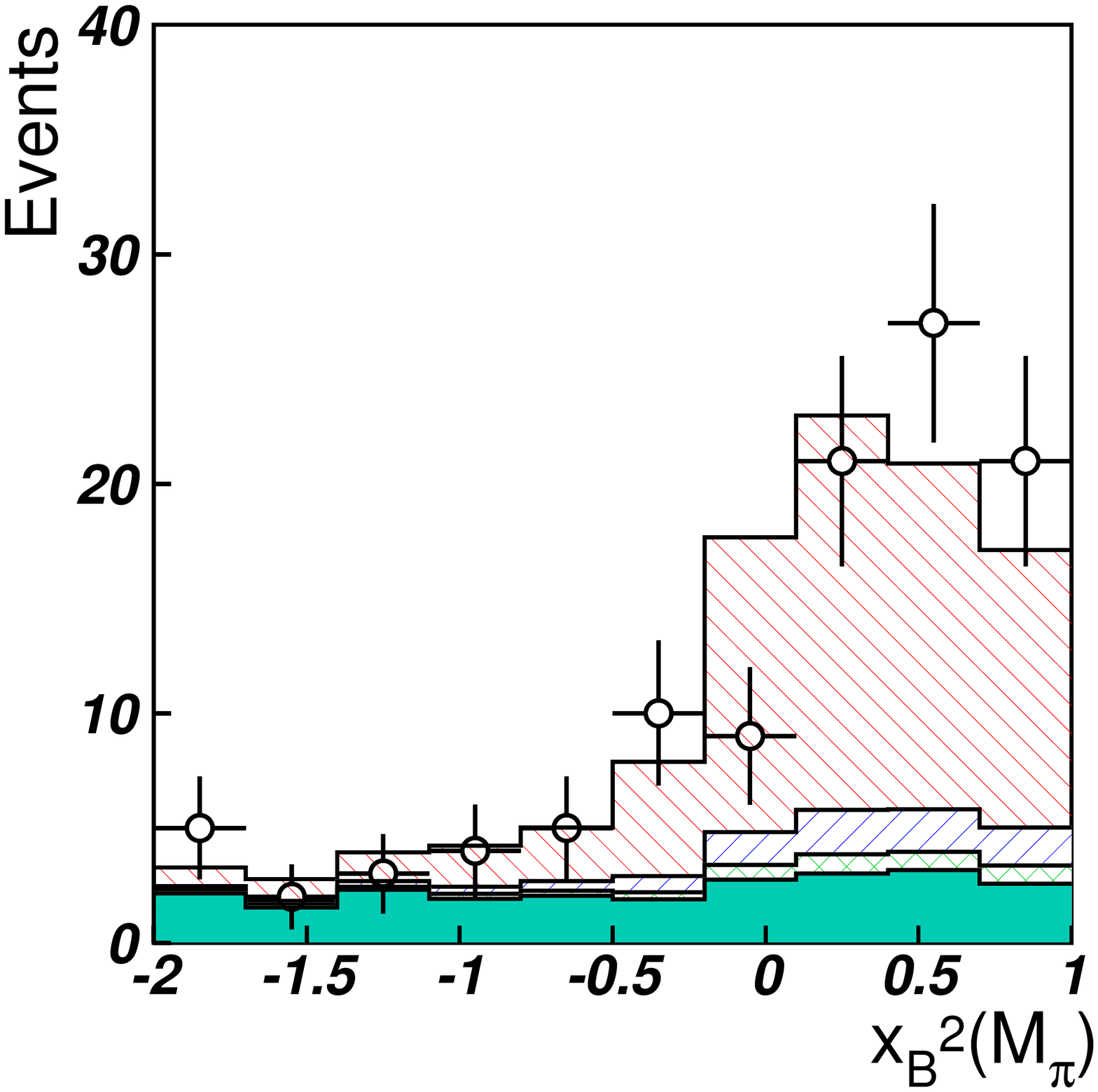,width=2.3in, height=2.3in, angle=0, scale=1.0 } }}
   \hspace{-0.8cm}{\mbox{\psfig{figure=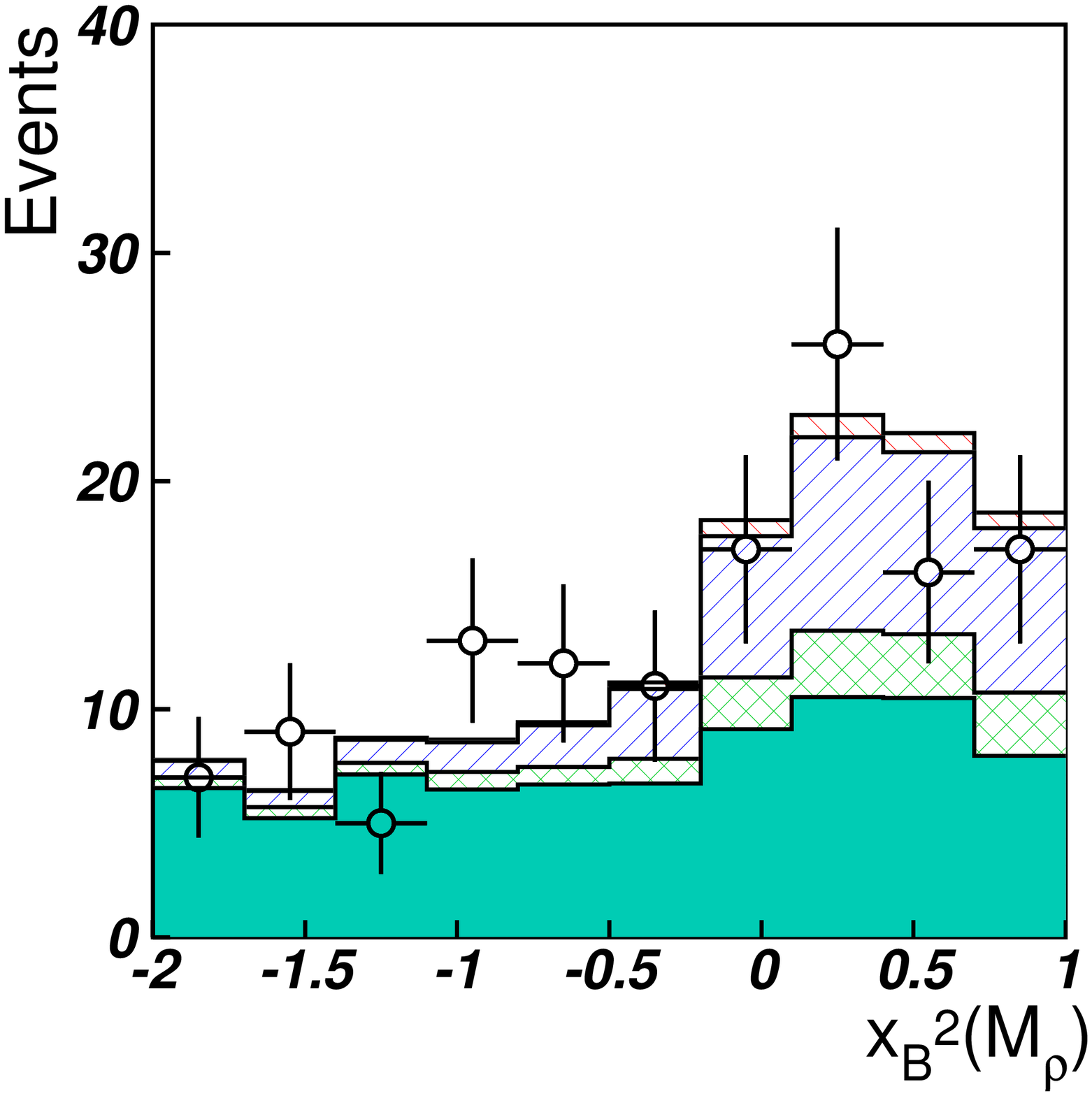,width=2.3in, height=2.3in, angle=0, scale=1.0 } }}
   \end{tabular}
  \caption{Projected $M_X$ distribution(left) and $x_B^2$ distributions 
  for the mass region of $\pi^-$ ($M_X< 0.18 $GeV/$c^2$, middle) 
  and $\rho^-$ ($ 0.53 < M_X < 1.1 $GeV/$c^2$, right) in all $q^2$ region; open circle is data.
  Histogram components are $\pi^- \ell^+ \nu$ (red $135^{\circ}$ hatch),
  $\rho^- \ell^+ \nu$ (blue $45^{\circ}$ hatch), other $X_u \ell^+ \nu$
  (green cross-hatch) and $B \bar{B}$ background (bluish green shaded).}
  \label{fig:fit_allq2}
 \end{center}
\end{figure}

Figure~\ref{fig:fit_allq2} presents the projection on $M_X$ and
$x_B^2$ of the fitting result for the data in the whole $q^2$ region.
The extracted yields for the signal components are 
$N(\pi^- \ell^+ \nu) = 72 \pm 11$ and 
$N(\rho^- \ell^+ \nu) = 59 \pm 15$,
with the LCSR model used for the two signal PDF.

%

\begin{figure}[htbp]
 \begin{center}
  \begin{tabular}{ccc}
   \hspace{-0.8cm}
   \mbox{\psfig{figure=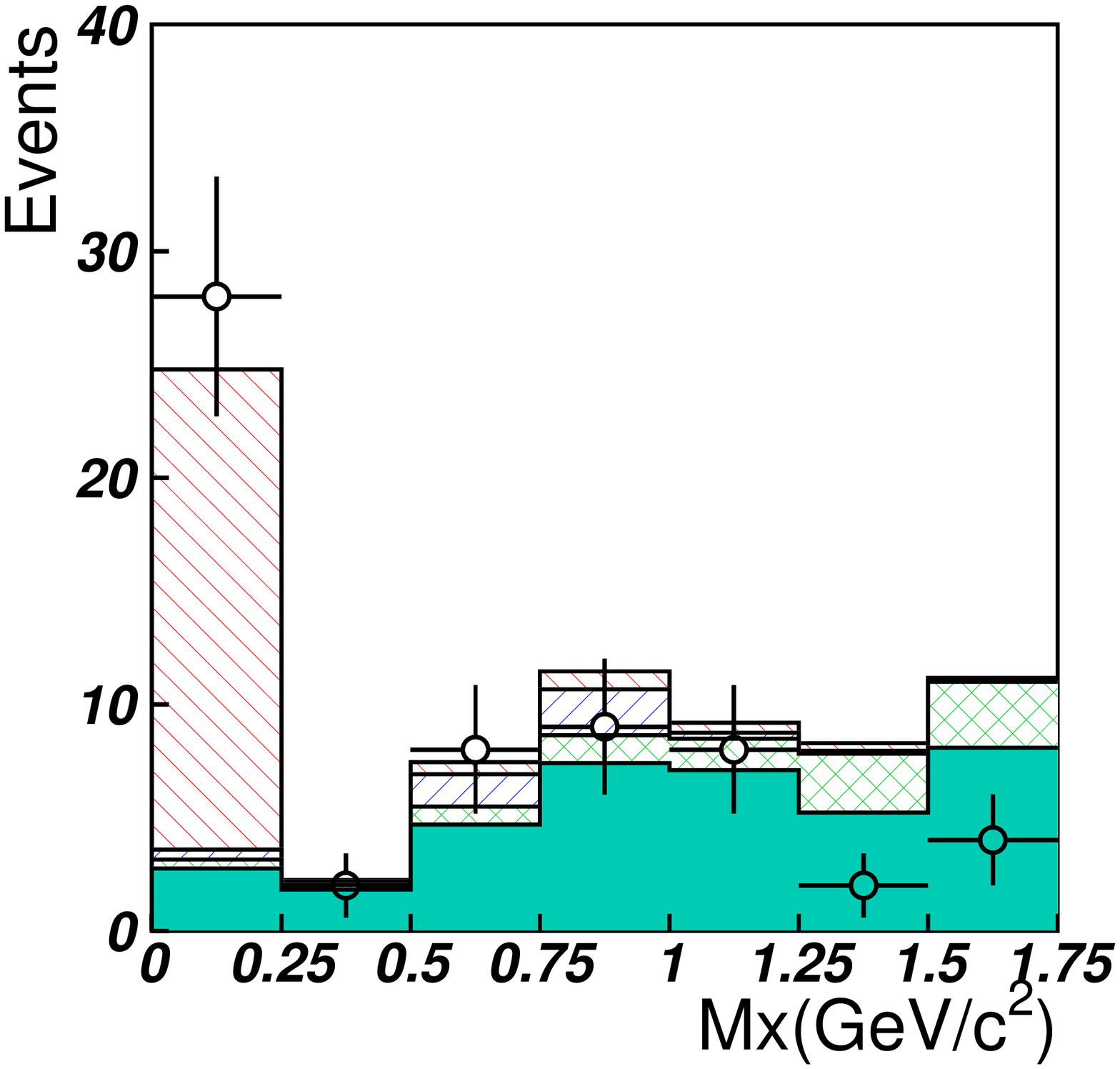,width=2.3in, height=2.3in, angle=0, scale=1.0 } } &
   \hspace{-0.8cm}{\mbox{\psfig{figure=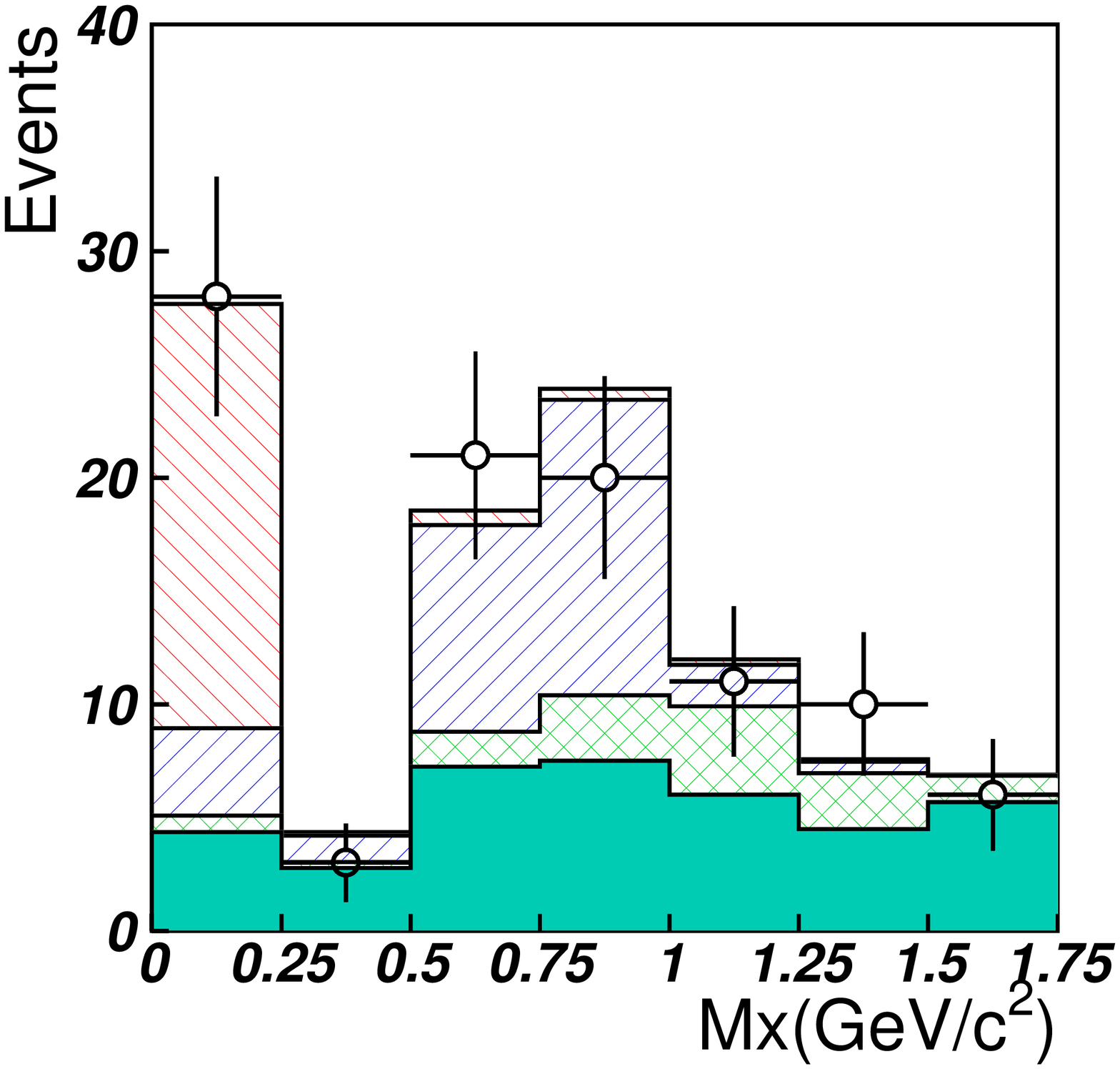,width=2.3in, height=2.3in, angle=0, scale=1.0 } }} &
   \hspace{-0.8cm}{\mbox{\psfig{figure=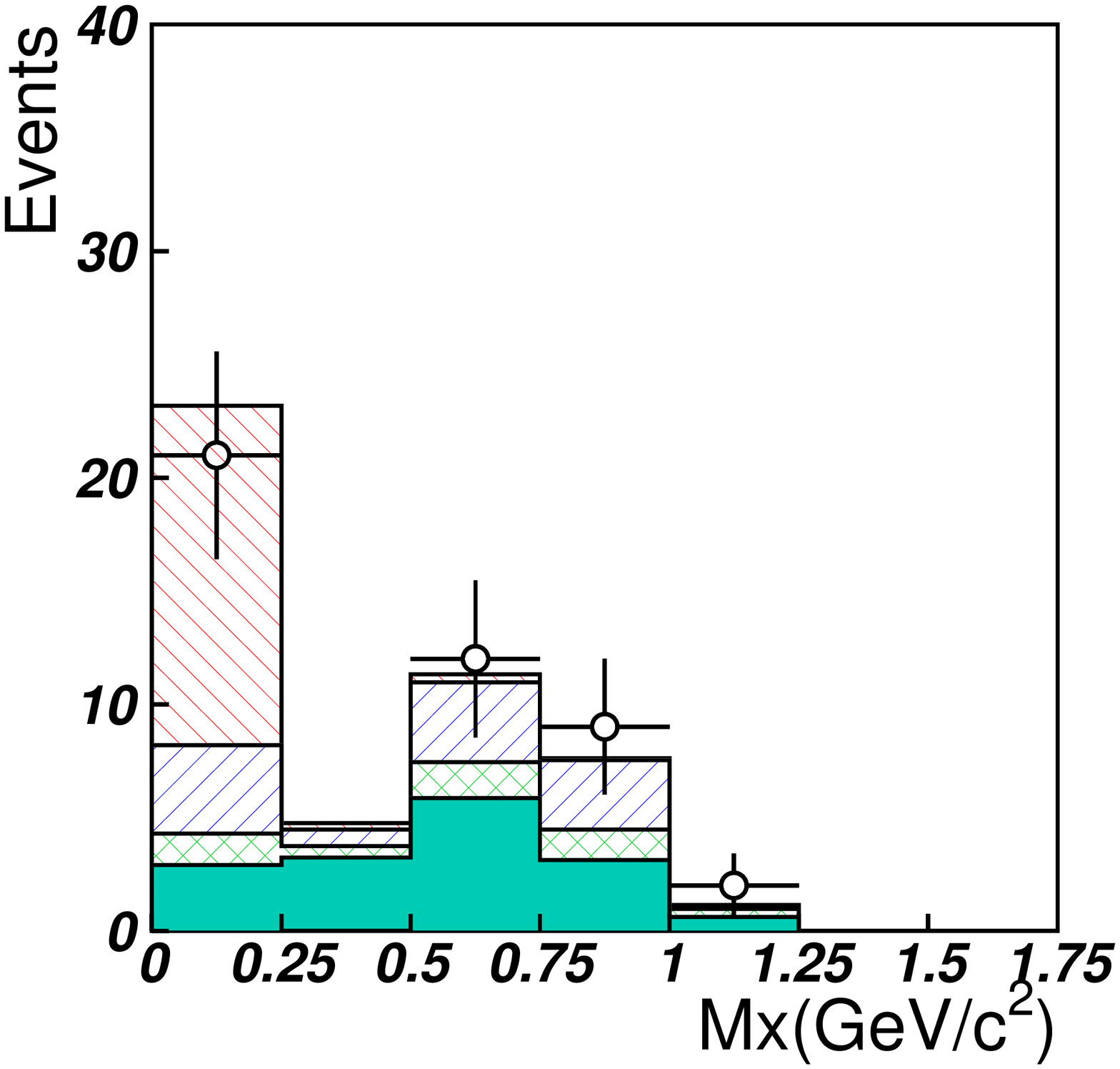,width=2.3in, height=2.3in, angle=0, scale=1.0 } }} \\
   \hspace{-1.0cm}{$q^{2} < 8$\,(GeV$^2/c^2$)}   & \hspace{-1.0cm}{$8 \le q^{2} < 16$ (GeV$^2/c^2$)}  & \hspace{-1.0cm}{$ q^{2} \ge 16$ (GeV$^2/c^2$)} \\
  \end{tabular}
  \caption{Projected $M_X$ distribution in each $q^2$ region;
  Open circle is data. Histogram components are $\pi^- \ell^+ \nu$ (red $135^{\circ}$ hatch),
  $\rho^- \ell^+ \nu$ (blue $45^{\circ}$ hatch), other $X_u \ell^+ \nu$
  (green cross-hatch) and $B \bar{B}$ background (bluish green shaded).}
  \label{fig:fit_subq2}
 \end{center}
\end{figure}

Figure~\ref{fig:fit_subq2} shows projections of the data, separated into three $q^2$ bins, 
$8 \leq q^2 < 16$ GeV$^2/c^2$ and $q^2 \geq 16$ GeV$^2/c^2$.
Here the normalization of the other $B^0 \to X_u^- \ell^+ \nu$
and the $B \bar{B}$ background components are fixed to those obtained in
the above fitting for the whole $q^2$ region.
Table~\ref{tbl:FFdep_q2_pilnu} and Table~\ref{tbl:FFdep_q2_rholnu}
summarize the extracted branching fractions with different FF-models
for the $\pi^- \ell^+ \nu$ and $\rho^- \ell^+ \nu$, respectively.
The results are unfolded using an efficiency matrix that relates 
the true and reconstructed $q^2$ for the three $q^2$ intervals. 
We calculate the total branching fraction by taking sum of the branching
fractions in the three $q^2$ intervals.

\begin{table}[htbp]
 \begin{center}
  \caption{Extracted branching fractions for $B^0 \to \pi^- \ell^+ \nu$ 
    with different FF models in unit of 10$^{-4}$; the total branching fraction and the 
    partial branching fractions in three $q^2$ intervals. 
    $\chi^2/n$ and the probability of $\chi^2$ shows the quality of the fit 
    of the FF shape to the extracted $q^2$ distribution.}
  \label{tbl:FFdep_q2_pilnu}
  \begin{tabular}{c|llllcc}
    \hline\hline
    FF model & ${\cal B}_{total}$ & ${\cal B}_{<8}$ & ${\cal B}_{8-16}$ & ${\cal B}_{\geq 16}$ 
                                                     & $\chi^2/n$  & {\it Prob.}\\ 
    \hline
    Ball'01  & $1.77 \pm 0.28$ & $0.72 \pm 0.15$ & $0.57 \pm 0.16$ & $0.48 \pm 0.17$ & 2.4/2 & 30.3\% \\
    ISGW~II  & $1.75 \pm 0.28$ & $0.71 \pm 0.15$ & $0.60 \pm 0.17$ & $0.44 \pm 0.16$ & 1.9/2 & 37.8\% \\
    UKQCD    & $1.75 \pm 0.27$ & $0.70 \pm 0.15$ & $0.60 \pm 0.16$ & $0.45 \pm 0.16$ & 0.4/2 & 81.0\% \\
    Average  & $1.76 \pm 0.28$ & $0.71 \pm 0.15$ & $0.59 \pm 0.16$ & $0.46 \pm 0.17$ &   --  &  --    \\
    \hline\hline
  \end{tabular}
 \end{center}
\end{table}

\begin{table}[htbp]
 \begin{center}
  \caption{Extracted branching fractions for $B^0 \to \rho^- \ell^+ \nu$ 
    with different FF models in unit of 10$^{-4}$; the total branching fraction and the 
    partial branching fractions in three $q^2$ intervals. 
    $\chi^2/n$ and the probability of $\chi^2$ shows the quality of the fit 
    of the FF shape to the extracted $q^2$ distribution.}
  \label{tbl:FFdep_q2_rholnu}
  \begin{tabular}{c|llllcc}
    \hline\hline
    FF model & ${\cal B}_{total}$ & ${\cal B}_{<8}$ & ${\cal B}_{8-16}$ & ${\cal B}_{\geq 16}$ 
                                                          & $\chi^2/n$  & \it{Prob.}\\ 
    \hline
    Ball'98 & $2.59 \pm 0.73$ & $0.24 \pm 0.35$ & $1.86 \pm 0.51$ & $0.50 \pm 0.39$ &  6.3/2 &  4.3\% \\
    ISGW~II & $2.50 \pm 0.86$ & $0.18 \pm 0.57$ & $1.76 \pm 0.48$ & $0.56 \pm 0.43$ &  1.3/2 & 52.2\% \\
    Melikhov& $2.79 \pm 0.83$ & $0.20 \pm 0.29$ & $1.88 \pm 0.51$ & $0.72 \pm 0.59$ &  8.2/2 & 1.6\% \\
    UKQCD   & $2.28 \pm 0.65$ & $0.22 \pm 0.37$ & $1.73 \pm 0.46$ & $0.33 \pm 0.27$ &  4.2/2 & 12.1\% \\
    Average & $2.54 \pm 0.78$ & $0.21 \pm 0.40$ & $1.81 \pm 0.49$ & $0.53 \pm 0.42$ &    --  &    --  \\
    \hline\hline
  \end{tabular}
 \end{center}
\end{table}

Table~\ref{tbl:systematic} summarizes the considered experimental systematic 
errors for the branching fractions.
The total systematic error is the quadratic sum of all individual ones.
A major contribution comes from the efficiency calibration with the
$B_{sig} \to D^{*-} \ell^+ \nu$ sample, where 8.3\% originates from
the statistics of the detected $D^{*-} \ell^+ \nu$ decays and 4.9\%
from the error on ${\cal B}(B^0 \to D^{*-} \ell^+ \nu)$ quoted in
~\cite{PDG2004}.
We consider the uncertainty in the number of $B^0 \bar{B^0}$ pairs; 
the ratio of $B^+B^-$ to $B^0 \bar{B^0}$ pairs ($f_+/f_0$,
2.4\%), the mixing parameter ($\chi_d$, 1.0\%) and the measured
number of $B \bar{B}$ pairs ($N_{B\bar{B}}$, 0.5\%).

The systematic error due to the uncertainty on the inclusive branching  
fraction ${\cal B}(B \to X_u \ell \nu)$, which is used to constrain 
$X_u^- \ell^+ \nu$ background, is estimated by varying this parameter
within $\pm 1 \sigma$ of the error.
The uncertainty on the $B \bar{B}$ background shape in our selection cut 
($N_{\pi^+} = 1, N_{\pi^0} \leq 1$) is studied in the simulation by 
randomly removing charged tracks and $\pi^0$ according to the error in detection 
efficiency (1\% for a charged track, 3\% for $\pi^0$),
and also by reassigning identified charged kaons as pions 
according to the uncertainty in the kaon identification efficiency (2\%).
The resultant change in the extracted branching fractions are assigned
as a systematic error. We have seen significant change for $\rho^- \ell^+ \nu$.


\begin{table}[htbp]
 \begin{center}
  \caption{Summary of systematic errors(\%) of the branching fractions}
  \label{tbl:systematic}
  \begin{tabular}{c|cc}
   \hline\hline
   Source & $\pi^{-} \ell^{+} \nu$ & $\rho^{-} \ell^{+} \nu$ \\
   \hline
   Tracking efficiency        &           1        &           1               \\ 
   $\pi^{0}$ reconstruction   &          --        &           3               \\ 
   Lepton identification      &         2.1        &         2.1               \\ 
   Kaon identification        &           2        &           2               \\ 
   $D^* \ell \nu$ calibration &         9.8        &         9.8               \\ 
   $Br(X_u \ell \nu)$ in the fitting        
                              &         0.2        &         3.4               \\ 
   $B\bar{B}$ background shape
                              &         4.4        &        31.5               \\
   $N_{B\overline{B}}$        &         0.5        &         0.5               \\ 
   $\it{f}_+ / \it{f}_0$      &         2.4        &         2.4               \\ 
   $\chi_d$                   &         1.0        &         1.0               \\ 
   \hline
   total                      &         11.5       &        33.5               \\ 
   \hline\hline
  \end{tabular}
 \end{center}
\end{table}

The FF model dependence of the extracted branching fractions has been 
studied by repeating the above fitting procedure by varying the FF 
model for the signal mode and also for the cross-feed mode 
($\pi \ell \nu \leftrightarrow \rho \ell \nu$).
For the extracted ${\cal B}(B^0 \to \pi^- \ell^+ \nu)$, the FF model 
dependence is found to be small; the variation from the average is 
$<0.6$\% for the change of the $\pi \ell \nu$ FF model and $<1.0$\% for 
the change of the $\rho \ell \nu$ FF mode.
As for ${\cal B}(B^0 \to \rho^- \ell^+ \nu)$, significant dependence on the
$\rho \ell \nu$ FF model is found; the variation is $<10.7$\% for the change 
of the $\rho \ell \nu$ FF model, while it is only $<1.3$\% for the change of 
the $\pi \ell \nu$ FF model.
A similar tendency was observed in the CLEO analysis~\cite{CLEO2003}.
At this stage, by adding linearly the maximum variations with the signal and 
cross-feed FF models, we conservatively assign 1.6\% (12\%) as the 
systematic error due to the FF model dependence for the 
$\pi^-(\rho^-) \ell^+ \nu$ total branching fraction.

\section{Results}
The total branching fractions are obtained by taking the simple 
average of the values obtained from  the FF models shown in Tables~\ref{tbl:FFdep_q2_pilnu} 
and ~\ref{tbl:FFdep_q2_rholnu}:
\begin{eqnarray}
{\cal B}(B^0 \to \pi^- \ell^+ \nu) & = &
(1.76 \pm 0.28 \pm 0.20 \pm 0.03) \times 10^{-4}, \\
{\cal B}(B^0 \to \rho^- \ell^+ \nu) & = &
(2.54 \pm 0.78 \pm 0.85 \pm 0.30) \times 10^{-4},
\end{eqnarray}  
where the errors are statistical, experimental systematic,
and systematic due to form-factor uncertainties.
The obtained branching fractions are consistent with the existing 
measurements by CLEO for  $B^0 \to \pi^-/\rho^- \ell^+ \nu$~\cite{CLEO2003} 
and BaBar for $B^0 \to \rho^- \ell^+ \nu$~\cite{BABAR2003}, within the 
measurement uncertainties.

\begin{figure}[htbp]
 \begin{center}
  \hspace{-0.5cm}{\mbox{\psfig{figure=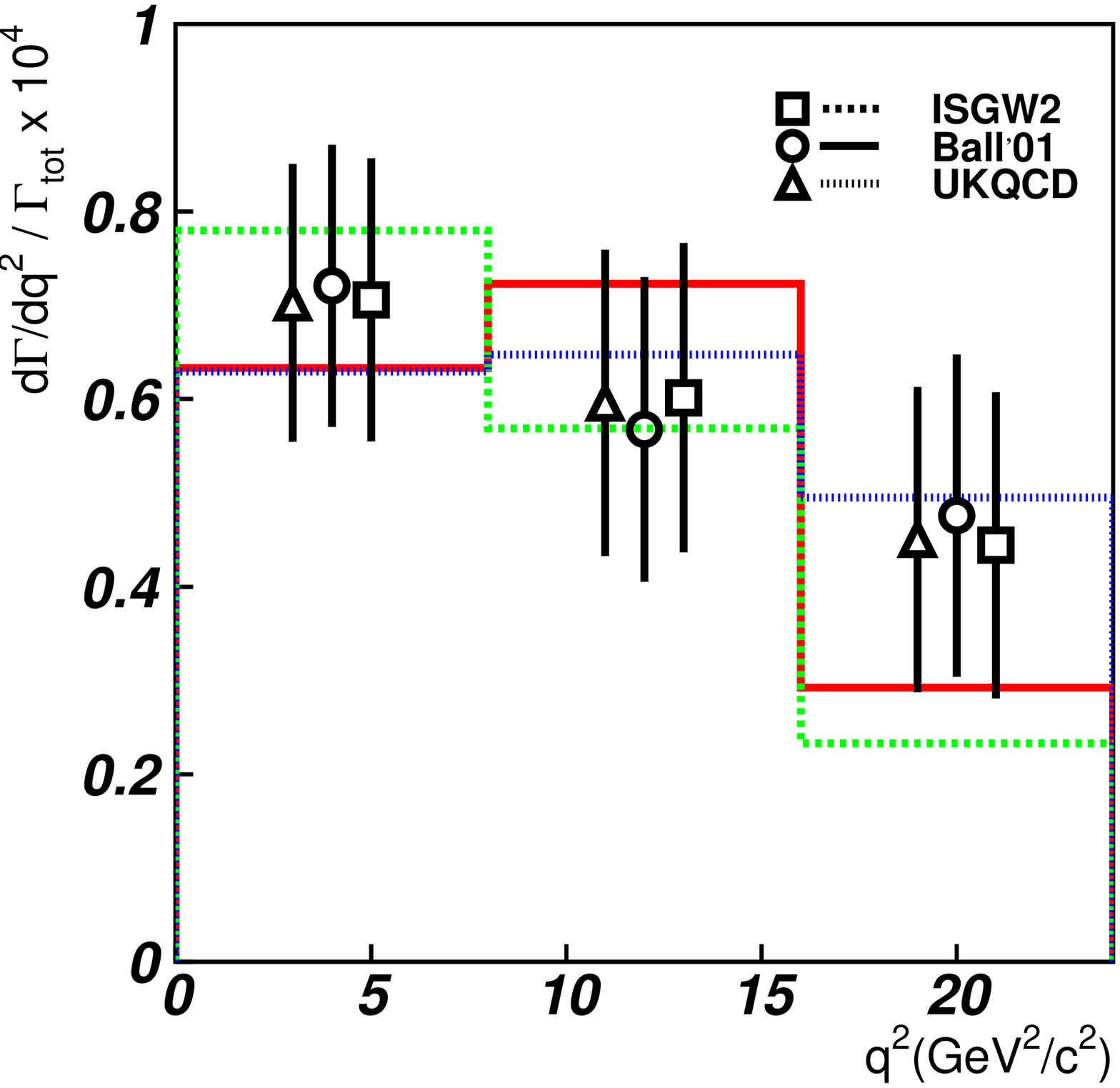,width=2.5in, height=2.5in, angle=0, scale=1.3 } } }
  \hspace{-0.5cm}{\mbox{\psfig{figure=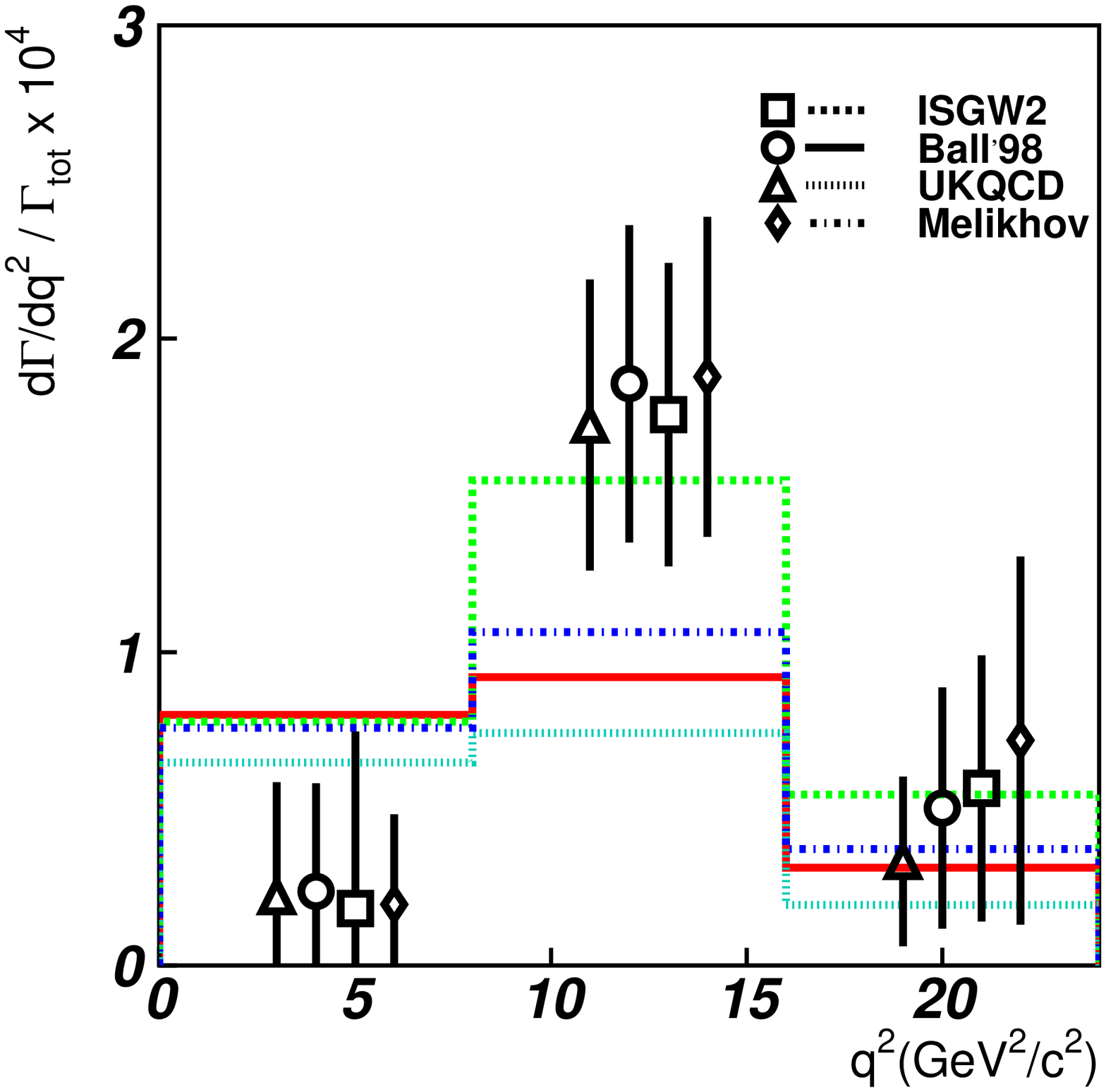,width=2.5in, height=2.5in, angle=0, scale=1.3 } }}
  \caption{Extracted $q^2$ distrubution for the $B^0 \to \pi^- \ell^+ \nu$(left) and
    $B^0 \to \rho^- \ell^+ \nu$(right) decays. Data points are shown for different FF
    models used to estimate the detection efficiency. 
    Lines are for the best fit of the FF shapes to the obtained $q^2$ 
    distribution.}
  \label{fig:q2dist}
 \end{center}
\end{figure}

Figure~\ref{fig:q2dist} presents the obtained $q^2$ distributions for the two
decay modes, overlaid with the best fits of FF shapes to the data.
To be self-consistent, the shape of a particular FF model is fit to the $q^2$ 
distribution extracted with that FF model.
The quality of the fit in terms of $\chi^2$ and the probability of $\chi^2$, 
shown in Table~\ref{tbl:FFdep_q2_pilnu} and 
~\ref{tbl:FFdep_q2_rholnu}, may provide one way to discriminate among the 
models.
At the present accuracy, we are unable to draw any conclusion on this point.

%
We extract $|V_{ub}|$ using the relation,
\begin{equation}
|V_{ub}| = \sqrt{ \frac{{\cal B}(B^0 \to \pi^-(\rho^-) \ell^+ \nu) }
  {{\tilde \Gamma}_{thy}~\tau_{B^0} } },
\end{equation}
where ${\tilde \Gamma}_{thy}$ is the form-factor normalization, predicted from theories.
In this paper, our major focus is on the $|V_{ub}|$ determination based 
on the $\pi^- \ell^+ \nu$ data and the form factor predicted by LQCD
calculations. 
Since the current LQCD calculations are available only in the region 
$q^2 \geq 16$ GeV$^2/c^2$, we use the branching fraction in the high $q^2$ bin 
extracted with UKQCD;
${\cal B}_{\geq 16} = (0.45 \pm 0.16) \times 10^{-4}$.
We use $\tau_{B^0} = 1.536 \pm 0.014$ ps for the $B^0$ lifetime
~\cite{PDG2004}.

\begin{figure}[htbp]
 \begin{center}
  \mbox{\psfig{figure=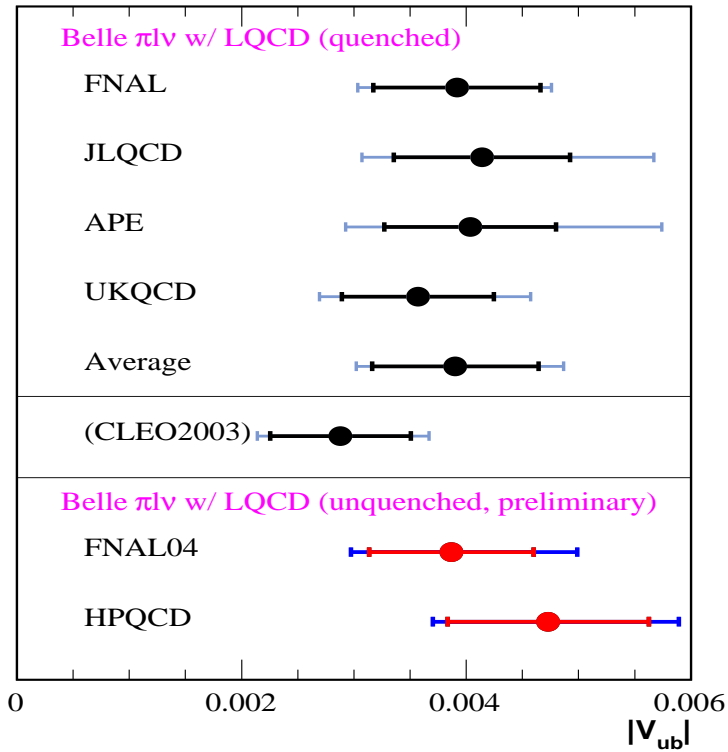,width=5.0in, height=5.0in, angle=0, scale=0.9 } } 
  \caption{ $|V_{ub}|$ determined using LQCD and the $\pi^- \ell^+ \nu$ 
    branching fraction in the $q^2 \geq 16$\,GeV$^2/c^2$ region.
    For each data point, $\pm 1 \sigma$ range of the experimental 
    errors (quadratic sum of statistical and systematic errors) 
    and the range of the total error including the theoretical error are indicated. 
    The CLEO result~\cite{CLEO2003} with the averaged quenched LQCD formfactor is also shown for comparison.}
  \label{fig:pilnu_lqcd}
 \end{center}
\end{figure}

We apply ${\tilde \Gamma}_{thy}$ predicted by the FNAL~\cite{FNAL}, 
JLQCD~\cite{JLQCD}, APE~\cite{CLEO2003} as well as UKQCD calculations,
as quoted by the CLEO analysis in 2003~\cite{CLEO2003}.
For the average of these results, the combined
${\tilde \Gamma}_{thy} = 1.92 ^{+0.32}_{-0.12} \pm 0.47$ ps$^{-1}$ 
calculated by CLEO work is used.
Here the errors are the statistical and the systematic in 
LQCD calculations, the latter including the quenching error of 15\%.
We obtain
\begin{equation}
|V_{ub}|_{(q^2 \geq 16)}^{\pi \ell \nu} =
(3.90 \pm 0.71 \pm 0.23 ^{+0.62}_{-0.48}) \times 10^{-3},  
\end{equation}
where the errors are statistical, experimental systematic, and theoretical.
Figure~\ref{fig:pilnu_lqcd} shows $|V_{ub}|$ determined with these
LQCD form factor predictions.
Our result is compatible and consistent with that from CLEO in 2003.

Recently, two preliminary results from the unquenched LQCD calculations were
reported, FNAL'04~\cite{Okamoto} and HPQCD~\cite{Shigemitsu}.
Their reported ${\tilde \Gamma}_{thy}$ for the $q^2 \geq 16$\,GeV$^2/c^2$ 
region are ${\tilde \Gamma}_{thy} = 1.96 \pm 0.51 \pm 0.39$ ps$^{-1}$ (FNAL'04)
and $1.31 \pm 0.13 \pm 0.30$ ps$^{-1}$ (HPQCD).
We also estimate $|V_{ub}|$ using these calculations:
\begin{eqnarray}
|V_{ub}|_{(q^2 \geq 16)}^{\pi \ell \nu} & = &
(3.87 \pm 0.70 \pm 0.22 ^{+0.85}_{-0.51}) \times 10^{-3}
~~~~~(\mbox{FNAL'04}), \\
|V_{ub}|_{(q^2 \geq 16)}^{\pi \ell \nu} & = &
(4.73 \pm 0.85 \pm 0.27 ^{+0.74}_{-0.50}) \times 10^{-3}
~~~~~(\mbox{HPQCD}).
\end{eqnarray}

\begin{figure}[htbp]
 \begin{center}
  \mbox{\psfig{figure=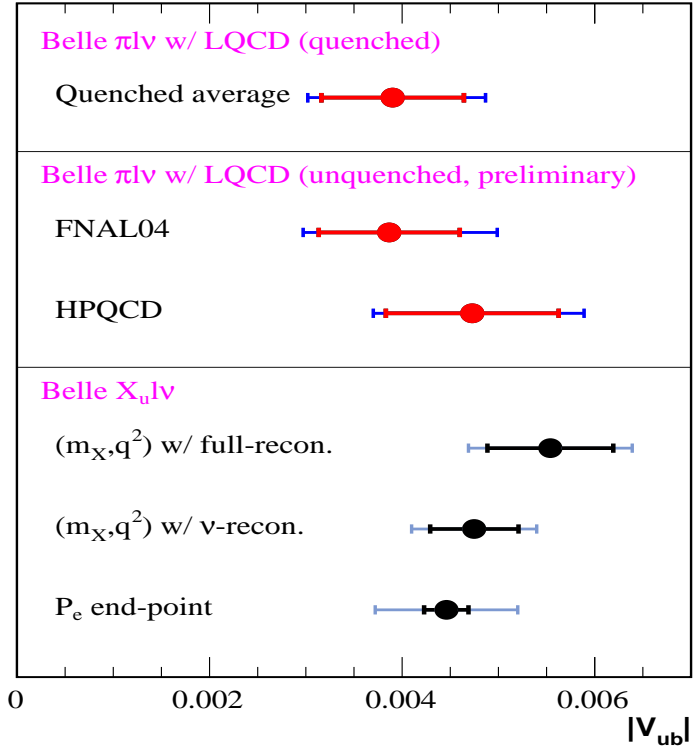,width=5.0in, height=5.0in, angle=0, scale=0.9 } } 
  \caption{ Summary of Belle $|V_{ub}|$ determination.
            Results based on the exclusive $B^0 \to \pi^- \ell^+ \nu$ measurement
            ($q^2 \geq 16$GeV$^2/c^2$) in this work with the averaged queched LQCD
            form factor (top) and with preliminary unqueched LQCD form factors
            (middle two) are shown, together with those based on our inclusive
            $B \to X_u \ell \nu$ measurements (bottom three). The lower two inclusive
             results are rescaled using the recent determination of the $f_u$ factor~\cite{limosani}.}
  \label{fig:belle_vub}
 \end{center}
\end{figure}

Finally, Figure~\ref{fig:belle_vub} summarizes the Belle $|V_{ub}|$ results 
from this work and from inclusive $B \to X_u \ell \nu$ measurements.
The above reported results from $B^0 \to \pi^- \ell^+ \nu$ high $q^2$ data
are in agreement with those from the inclusive $B \to X_u \ell \nu$ decays
~\cite{BELLE_Ilija, BELLE_Kakuno, BELLE_Limosani}.
In the future, by accumulating more luminosity and with improvement in
unquenched LQCD calculations, the present measurement will provide 
a more accurate determination of $|V_{ub}|$.

\vspace{1.0cm}

We thank M.Okamoto, J.Shigemitsu, S.Hashimoto and T.Onogi for
useful discussions of recent lattice QCD calculations.
We thank the KEKB group for the excellent operation of the
accelerator, the KEK Cryogenics group for the efficient
operation of the solenoid, and the KEK computer group and
the National Institute of Informatics for valuable computing
and Super-SINET network support. We acknowledge support from
the Ministry of Education, Culture, Sports, Science, and
Technology of Japan and the Japan Society for the Promotion
of Science; the Australian Research Council and the
Australian Department of Education, Science and Training;
the National Science Foundation of China under contract
No.~10175071; the Department of Science and Technology of
India; the BK21 program of the Ministry of Education of
Korea and the CHEP SRC program of the Korea Science and
Engineering Foundation; the Polish State Committee for
Scientific Research under contract No.~2P03B 01324; the
Ministry of Science and Technology of the Russian
Federation; the Ministry of Education, Science and Sport of
the Republic of Slovenia; the National Science Council and
the Ministry of Education of Taiwan; and the U.S.\
Department of Energy.


%

\end{document}